\documentclass[prb, aps, twocolumn, superscriptaddress,showpacs]{revtex4}
\usepackage[english]{babel} 
\usepackage[english]{layout}
\usepackage{amsmath,amsfonts,amssymb}
\usepackage{graphicx}
\usepackage{float}
\usepackage{color}
\usepackage{braket}
\usepackage{version}
\usepackage{array,multirow,makecell}
\usepackage[T1]{fontenc}
 
\begin{document}

\title{Angular-dependent Andreev reflection on a polaritonic superfluid}

\author{I. Septembre}
\affiliation{Universit\'e Clermont Auvergne, Clermont Auvergne INP, CNRS, Institut Pascal, F-63000 Clermont-Ferrand, France} 
\author{D. D. Solnyshkov}
\affiliation{Universit\'e Clermont Auvergne, Clermont Auvergne INP, CNRS, Institut Pascal, F-63000 Clermont-Ferrand, France} 
\affiliation{Institut Universitaire de France (IUF), 75231 Paris, France}
\author{G. Malpuech}
\affiliation{Universit\'e Clermont Auvergne, Clermont Auvergne INP, CNRS, Institut Pascal, F-63000 Clermont-Ferrand, France}

\begin{abstract}
We study analytically an analog of the Andreev reflection at a normal-superfluid interface. The polariton gapped superfluid region is achieved by quasi-resonant optical pumping. The interacting polaritons are described with the driven-dissipative Gross-Pitaevskii equation. We find analytical formulas for the angles and amplitudes of the reflected and transmitted particles. There are limit angles and energies, above which Andreev reflection/transmission cannot be observed anymore and where the Andreev wave becomes a surface mode, exponentially localized on the interface. These properties are confirmed by solving numerically the Gross-Pitaevskii equation in simulations reproducing realistic experimental conditions.
\end{abstract}
\maketitle

\section{Introduction}
Andreev reflection is an anomalous reflection occurring at the interface between a superconducting and a non-superconducting material~\cite{andreev1964thermal}. An electron, instead of being reflected with just its direction changed, is reflected as a hole excitation. Its energy is inverted with respect to the Fermi energy, and so are its charge, wave vector, and spin. Inversely, an incoming hole is reflected as an electron. Andreev reflection determines the properties of metal/superconductor structures~\cite{blonder1982transition,pannetier2000andreev}. It has been observed experimentally not only in electronic systems~\cite{bozhko1982observation,benistant1983direct} but also in superfluid helium experiments~\cite{enrico1993direct}. However, it has never been observed in fermionic gases despite a proposal~\cite{van2007normal}. The Andreev process can also lead to specular reflection when dealing with relativistic electron dispersions as in graphene~\cite{beenakker2006specular}.

A similar effect has been discovered and studied in the 70s in nonlinear optics. It is known as optical phase conjugation~\cite{yariv1978phase}. In this context, an incident optical wave arriving on a phase-conjugating mirror (which involves nonlinear processes) is reflected as its time-reversed partner, meaning that the reflected wave has the same frequency and opposite wave vector. The analogy between optical phase conjugation achieved using degenerate four-wave mixing and the electronic Andreev reflection has been noticed and discussed in Refs.~\cite{van1991andreev,PhysRevA.56.4216}.
Andreev reflection on a Bose-Einstein condensate has been proposed theoretically as well~\cite{zapata2009andreev} but has not been observed yet. In Ref.~\cite{zapata2009andreev}, the Bose-Einstein condensate was in a supersonic regime. This allows one to draw a parallel with phenomena occurring close to the horizon of a black hole. It is now understood that Andreev reflection is very similar to phenomena encountered near the horizon of black holes~\cite{PhysRevD.53.7082,PhysRevB.100.245436,PhysRevD.96.124011,PhysRevD.102.064028}, such as Hawking emission. Analog physics allows investigating the properties of inaccessible systems such as black holes in this case \cite{Unruh1981,jacquet2020polariton}.

Exciton-polaritons are quasi-particles arising from the strong coupling of excitons and cavity photons~\cite{kavokin2003cavity}. They exhibit strong interactions, and thus a polariton mode behaves as a nonlinear oscillator exhibiting bistability \cite{Baas2004}. Excitons-polaritons are part-light part-matter bosonic quasiparticles arising from the strong coupling between semiconductor excitons and confined photons~\cite{hopfield1958theory,kavokin2017microcavities}. 
They are intrinsically strongly interacting and their bosonic non-linearity can show up as a parametric oscillator behavior~\cite{savvidis2000angle,baumberg2000parametric,savvidis2001off,saba2001high,kundermann2003coherent,Demenev2008,septembre2021parametric,solnyshkov2021microcavity} or Bose-Einstein condensation~\cite{kasprzak2006bose,deng2010exciton,byrnes2014exciton,kavokin2022polariton,PhysRevLett.129.066802} even at room temperature~\cite{Baumberg2008,Feng2013,plumhof2014room,dusel2020room,tang2021room}. Under resonant pumping, the spectrum of exciton-polaritons modes elementary excitations above the bistability threshold can contain a gap~\cite{carusotto2004probing,ciuti2005quantum,Carusotto2013,kavokin2017microcavities}. This regime is called a gapped superfluid and is analogous to a superconductor. In previous work, we have considered the reflection of a plane wave at an energy $E+E_p$ on such superfluid of energy $E_p$ and phase $\phi$ ~\cite{septembre2021parametric}. We have shown that normal elastic reflection occurs together with an Andreev reflection analog at an energy $E_p-E$ and with a phase shift $2\phi$. We have shown that this phase shift can be interpreted as an artificial gauge field allowing the implementation of topological bands in multi-terminal Josephson junctions~\cite{septembre2022weyl}.

In this work, we go further and study comprehensively the angular dependence of the Andreev reflection on a polaritonic gapped superfluid in 2 dimensions. We derive the Bogoliubov-de Gennes equations from the 2D Gross-Pitaevskii equation for a 2D planar microcavity hosting exciton-polaritons with quasi-resonant pumping. The interface between a non-superfluid region (not pumped) and a gapped superfluid region (pumped) is the theatre of an anomalous reflection, similar to the Andreev reflection, and a non-trivial transmission (for energies of incident particles larger than the gap). We solve the problem analytically, finding the different quantities characterizing the reflection and the transmission, such as wavevectors and decay lengths, angles, critical angles and energies, and the scattering coefficients. We moreover find peculiar solutions that take the form of surface modes for the Andreev wave. The analytical results are verified by solving numerically the 2D Gross-Pitaevskii equations. These results can be useful to study polaritonic multi-terminal Josephson junctions that host Weyl singularities~\cite{septembre2022weyl}, and also in the study of phenomena occurring close to the horizon of black holes.


\section{Gapped superfluid regime}
\begin{figure}[tbp]
    \centering
    \includegraphics[width=0.99\linewidth]{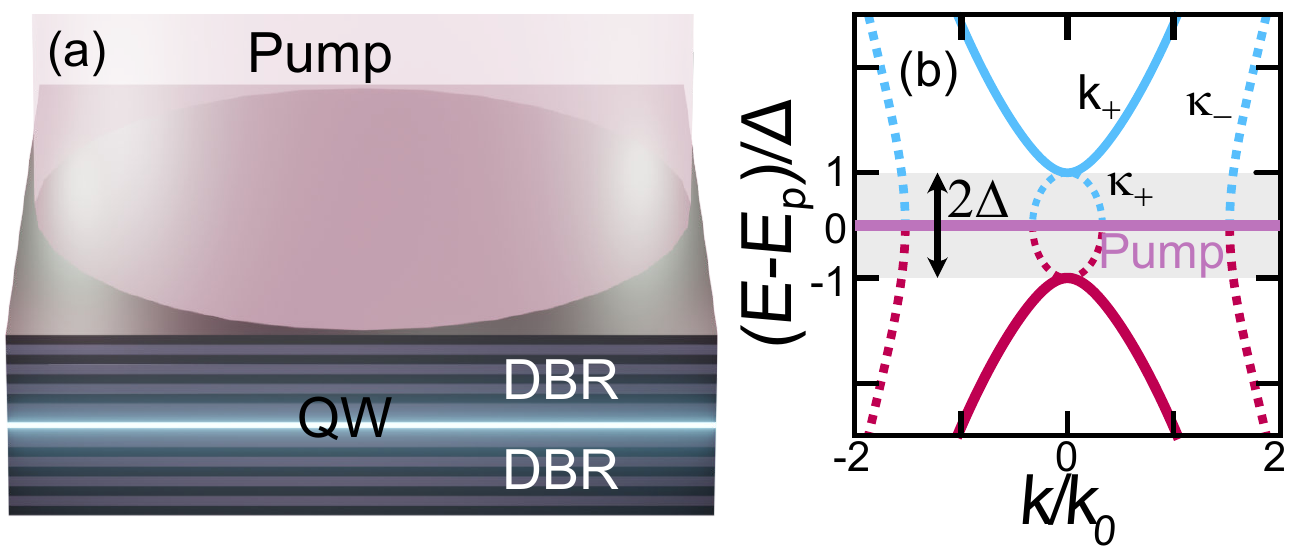}
    \caption{(a) Sketch of the planar optical microcavity hosting exciton-polaritons. light is confined in the $z$ direction by distributed Bragg reflectors (DBR), while excitons are confined by a quantum well (QW). The external pumping is provided by a laser at energy $E_p$. (b) Energy dispersion of the propagative states (solid lines) associated with the wave vector $k_+$ and energy dependence of the inverse decay lengths (dashed lines) $\kappa_\pm$ in the gapped superfluid regime. The gap is the shaded area where no propagative states exist, and is centered around the pump energy (purple line). Blue - positive energy $+E$; red - negative energy $-E$. $k_0=\sqrt{2 m E_p}/\hbar$.}
    \label{fig1}
\end{figure}

In this section, we consider a planar microcavity hosting interacting photons (exciton-polaritons) under quasi-resonant pumping. We neglect polarisation effects. We remind the existence of a gapped superfluid regime and the subsequent elementary excitations of the system (bogolons). An originality here is that we consider both the well-known propagative states and the spatially decaying evanescent solutions which, in particular, exist in the gap and which are important to describe the reflection processes at the gap energies. 
We study a strongly-coupled planar microcavity~\cite{Carusotto2013,goblot2016phase,kavokin2017microcavities} composed of two distributed Bragg reflectors surrounding a quantum well, as depicted in Fig.~\ref{fig1}(a).

This type of cavity allows confining exciton-polaritons in the $z$ direction, letting them propagate in the $(x,y)$ plane where they can be described by a parabolic dispersion. We first consider the situation where the whole cavity is under quasi-resonant pumping. Experimentally, a pump laser drives the whole cavity from the top, as shown in Fig.~\ref{fig1}(a).
The wave function $\psi$ of a quasiparticle confined in the cavity can be described using the 2D Gross-Pitaevskii equation:
\begin{equation}\label{GP}
    i\hbar \frac{\partial\psi}{\partial t}=\left[ -\frac{\hbar^2}{2m}\bm{\nabla}^2 -i\gamma +\alpha |\psi|^2 \right] \psi +P,
\end{equation}
where $\hbar$ is the reduced Planck constant, $m$ the effective mass of the exciton-polaritons, $\gamma$ their lifetime, $\alpha>0$ the repulsive interaction between them, and $P=e^{-i\omega_p t}$ the pumping term at the energy $E_p=\hbar \omega_p$. We consider $\gamma \to 0$ to simplify the analytics, but we verify the validity of the solutions obtained with this assumption later. We neglect polarization effects.
This equation has a spatially homogeneous solution $\psi_s=\sqrt{n}e^{i\phi}e^{-i\omega_p t}$ demonstrating a bistable behavior~\cite{PhysRevA.69.023809}. The density of exciton-polaritons $|\psi|^2=n$ can be directly measured experimentally because it dictates the intensity of the light emitted by the cavity.
Additionally to the spatially homogeneous solution, we consider the weak superfluid excitations called bogolons which result from the non-linear coupling between two complex conjugate particles:
\begin{equation}\label{wf}
    \psi=e^{-i\omega_p t}\left ( \psi_s+u e^{i\bm{k} \bm{r}}e^{-i\omega t}+v^* e^{-i \bm{k} \bm{r}}e^{i\omega^* t}\right ),
\end{equation}
where $u,~v$ are the Bogoliubov coefficients. We can obtain the Bogoliubov-de Gennes equations by inserting the wave function~\eqref{wf} into the Gross-Pitaevskii equation~\eqref{GP}:
\begin{equation}\label{BdG0}
\begin{pmatrix}
 \mathcal{L} && \alpha \psi_s^2 \\
 -\alpha \psi_s^{*2}  && -\mathcal{L^*}
\end{pmatrix}\begin{pmatrix}
u\\v
\end{pmatrix}=E\begin{pmatrix}
u\\v
\end{pmatrix},
\end{equation}
where $\mathcal{L}=\left(\epsilon_\mathbf{k} -E_p+2\alpha n\right)$ with $\epsilon_k=\hbar^2 k^2/2m$ ($|\mathbf{k}|=k$). $E=\hbar\omega$ denotes the energy of a particle with respect to the pump energy $E_p$.
The energy $E$ comes straightforwardly from Eq.~\eqref{BdG0}:
\begin{equation}\label{disp}
    E^2=\left( \epsilon_\mathbf{k}+\alpha n -E_p\right)\left( \epsilon_\mathbf{k}+ 3\alpha n -E_p\right).
\end{equation}

One can use Eq.~\eqref{disp} to determine the wave vectors of the particles depending on their energy:
\begin{equation}\label{kpm}
    k_\pm = \frac{\sqrt{2m \left( E_p - 2 \alpha n \pm \sqrt{(\alpha n)^2+E^2} \right)}} { \hbar}.
\end{equation}
If we only consider the case of repulsive interactions $\alpha n>0$, there are four different configurations of the system. First, the superfluid regime is obtained when $\alpha n\geq E_p$. The limit case $\alpha n=E_p$ gives the linear Bogoliubov spectrum, equivalent to the one found in the case of thermal equilibrium $\mu=\alpha n$, which is $E^2=\epsilon_k(\epsilon_k+2\alpha n)$. Then, for $\alpha n>E_p$, the superfluid is gapped, meaning that the two branches of the dispersion are separated by a gap $2\Delta$, centered at the pump energy, with
\begin{equation}\label{gap}
    \Delta=\sqrt{(\alpha n-E_p)(3\alpha n-E_p)}.
\end{equation}

In addition to these two superfluid regimes, where one can find the dispersion of excitations with completely real energies, there are different cases for $E_p>\alpha n>0$ where the dispersion is complex, including rings of exceptional points, a non-Hermitian degeneracy line~\cite{Voigt1902,zhen2015spawning,cerjan2019experimental,PhysRevLett.123.227401,mc2020weyl,li2021experimental,krol2022annihilation}. We will not consider these configurations in the following. We study only the gapped superfluid configuration where $\alpha n>E_p$. We consider not only propagative states, but also evanescent solutions (with imaginary wave vector).
Indeed, from Eq.~\ref{kpm}, we see that the condition $\alpha n> E_p$ we set gives imaginary wave vectors for certain energies, that is, evanescent states. More precisely, the wave vector $k_-$ is imaginary for all energies  and the wave vector $k_+$ is real only for $E>\Delta$. This means that instead of the wave vectors~\eqref{kpm}, one should use inverse decay lengths:
\begin{equation}\label{kkpm}
    \kappa_\pm = \sqrt{ \frac{2m \left(2\alpha n-E_p\mp\sqrt{(\alpha n)^2+E^2} \right)} { \hbar^2}},
\end{equation}
which are the direct extensions of $k_\pm$, respectively, to the case of imaginary wave vectors. 

We plot the dispersion of the propagative states and the dependence of the inverse decay lengths on the energy in Fig.~\ref{fig1}(b). The dispersion obtained from Eq.~\eqref{disp} shows propagative states (solid lines) that correspond to real energies and demonstrate a gap. There are also evanescent states, associated with $\kappa_+$ (for $E<\Delta$) and $\kappa_-$ (for all energies). The dispersion of these states is shown in dashed lines in Fig.~\ref{fig1}(b). When the energy of a particle lies inside the gap $E<\Delta$, there are only evanescent states in the superfluid. However, we insist that the evanescent states are present in the superfluid even outside the gap, together with the propagative states: they are important to describe the interfaces in finite-size systems, as we will see below.

The parts of the wave function oscillating at frequencies $\pm E/\hbar$ are determined by the Bogoliubov coefficients $u,~v$.
They can be found from Eq.~\eqref{BdG0} with an appropriate normalization condition. Two different normalization conditions can be used. A bogolon is a particle of energy $E$ (which can be positive or negative) with fractions $|u|^2$ at $E$ and $|v|^2$ at $-E$. For a positive energy $E$, we have $E|u|^2-E|v|^2=1$, whereas for a negative energy $E$, we have $E|u|^2-E|v|^2=-1$. The first possibility is to set $|u|^2-|v|^2=1$ for both positive and negative energies, and set $E>0$, so that a bogolon of negative energy will be at the energy $-E$, leading to $E|u|^2-E|v|^2=-1$. We do not choose this condition. We rather choose that a bogolon is a particle of energy $E$ which can be positive or negative, and use the normalization condition:
\begin{equation}\label{norm}
    \begin{array}{r c l}
      |u|^2-|v|^2 = 1~&\mathrm{for}&~E>0,\\
      |u|^2-|v|^2 = -1~&\mathrm{for}&~E<0.
   \end{array}
\end{equation}
This means that in the case $E>0$, the amplitude of the mode at $+E$ denoted $|u|^2$ will be larger than the amplitude at $-E$ which is $|v|^2$ ($|u|>|v|$) and \textit{vice versa} for a negative energy, which is self-consistent:
\begin{equation}
\begin{array}{r c l}
    E>0 \implies |u|>|v|,\\
    E<0 \implies |u|<|v|.
    \end{array}
\end{equation}

The analytical formulas of the coefficients can be computed from Eqs.~\eqref{BdG0} and~\eqref{norm}:
\begin{equation}
    \begin{array}{r c l}\label{u-v}
      u_\pm & = & \frac{\sqrt{\sqrt{(\alpha n)^2+E^2}\mp E}}{\sqrt{2E}}e^{i\phi},\\
      v_\pm & = &\pm \frac{\sqrt{\sqrt{(\alpha n)^2+E^2}\pm E}}{\sqrt{2E}}e^{-i\phi},
   \end{array}
\end{equation}
Note that the phase of the superfluid $\phi$ appears in the Bogoliubov coefficients with opposite signs (positive in $u$ and negative in $v$), which can be traced back to the conjugation in Eq.~\eqref{wf}.

Now that we have provided a detailed study of the different regimes achievable for a collective excitation of the gapped superfluid formed by quasi-resonant pumping of an optical microcavity, we can proceed to the description of systems containing interfaces.


\section{Andreev reflection analogue}
\begin{figure}[tbp]
    \centering
    \includegraphics[width=0.99\linewidth]{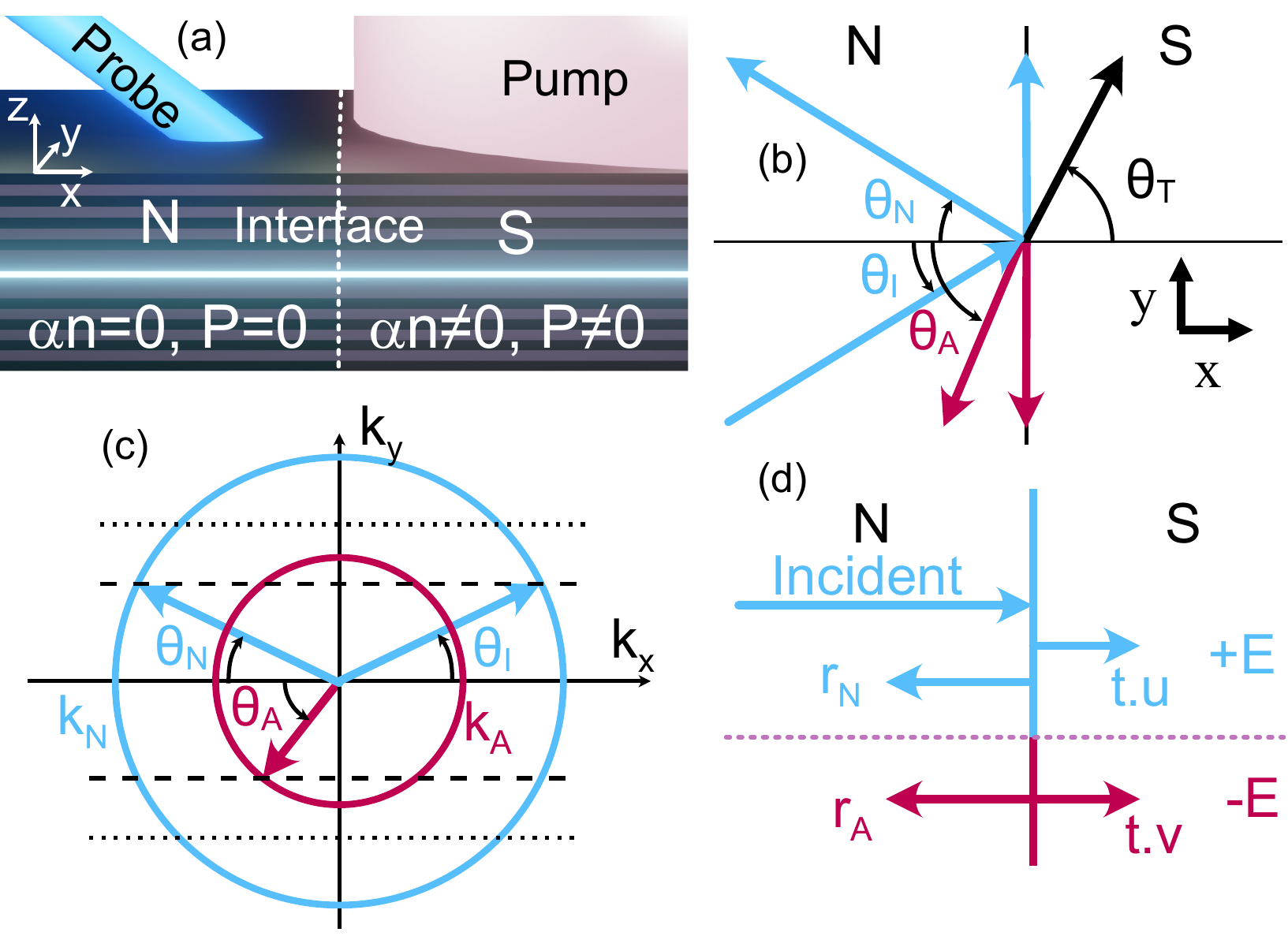}
    \caption{(a) Scheme of the cavity hosting exciton-polaritons, partly under resonant pumping creating a superfluid (red) region. A probe (blue) is sent from the normal region toward the interface (white dotted line) to observe Andreev reflection. (b) Scheme of the 2D real space for an incident wave at positive energy. The arrows denote the wave vectors with angle $\theta_{I,N,A,T}$ (with respect to the normal incidence) denoting incident, normally reflected, Andreev reflected, and transmitted wave at positive (blue) and negative (red) energies. (c) Scheme of the $k$-space where the norms $k_{N,A}$ appear as circles, the radius of the latter being smaller. This allows finding the angles because of the constant $k_y$ (dashed/dotted black lines). For a given $k_y$ (set by $E$ and $\theta_I$), Andreev reflection can be possible (dashed) or impossible (dotted black lines). (d) Energy representation of the scattering phenomena. A wave incident at positive energy is reflected both at the same energy and at its opposite (and is transmitted at both energies if $E>\Delta$).}
    \label{fig2}
\end{figure}

We now consider a different configuration where only half of the cavity is pumped, as shown in Fig.~\ref{fig2}(a). This creates an interface between the area under pumping and the area without pumping, that is, a normal-superfluid interface. We show that a phenomenon analogous to Andreev reflection can occur at this interface, and study its properties.

In our model, the interaction and pumping terms present a step-like profile, as shown in Fig.~\ref{fig2}(a). The interactions and pumping are null for $x<0$ and constant for $x>0$, which sets a sharp interface at $x=0$ between the two regions. The $x>0$ region is in the gapped superfluid regime $\alpha n>E_p$. For the normal region $x>0$, where there are no interactions nor pumping, Eq.~\eqref{GP} reduces to the 2D time-dependent Schr\"odinger equation:
\begin{equation}\label{sch}
    i\hbar \frac{\partial\psi}{\partial t}= -\frac{\hbar^2}{2m}\bm{\nabla}^2\psi.
\end{equation}
with plane waves solutions of the form:
\begin{equation}\label{wfnormal}
    \psi(x<0,y,t)=e^{\pm i \bm{k}.\bm{r}}e^{-i \epsilon_k t/\hbar}
\end{equation}
with:
\begin{equation}
    \epsilon_k=\frac{\hbar^2k^2}{2m}.
\end{equation}

On the contrary, on the right-hand side, the complete form of Eq.~\eqref{GP} has to be used. We will consider the problem of a wave incident from the normal region on the superfluid region.
However, depending on whether the particle has its energy inside or outside the gap, the wave function ansatz will be different, describing evanescent or propagative states in the superfluid. We first consider the case of energies of incident particles lying inside the gap $E<\Delta$, so that there can be no propagation in the superfluid, and therefore there is no transmission. This is a non-trivial extension of the previous 1D study  in~\cite{septembre2021parametric} to two dimensions. Then we consider energies outside the gap, leading to propagative states in the superfluid region.

\subsection{$E<\Delta$}
We consider the problem of an incident particle of energy $E<\Delta$ (the derivation made for an incident particle of energy $-E$ follows the same lines). Because we measure energies with respect to the pump energy, the norm of the wave vector is:
\begin{equation}
    k_N=\frac{\sqrt{2m(E_p+E)}}{\hbar}.
\end{equation}
If the incident wave has an angle of incidence $\theta_I$ ($0\leq\theta_I<\pi/2$) from the normal region toward the superfluid one, its wave vector is:
\begin{equation}
    \mathbf{k}_I=k_N\left(\cos\theta_I\mathbf{x}+ \sin\theta_I\mathbf{y}\right).
\end{equation} 
The wave vector is thus determined completely by the angle of incidence and the energy of the incident particles, that is, by two parameters than can be easily tuned in experiments with polaritons. 
\begin{equation}
    k_y=k_N\sin \theta_I,
\end{equation}
The invariance of the problem in the $y$ direction imposes that at a given energy the wave vector along $y$ is the same in the normal region and the superfluid. All waves at the energy $E$ of the incident wave possess the same $y$ component of the wave vector $k_y$. At the energy of the Andreev reflection, symmetric to $E$ with respect to $E_p$, the $y$ component of the wave vector is $-k_y$ because of the complex conjugation resulting from the parametric processes creating amplitudes at the Andreev energy.  
To use the scattering formalism, we need to define the different scattering states we will consider. First, we consider the incident wave
\begin{equation}
    \psi_I = \begin{pmatrix}1\\0\end{pmatrix} e^{i k_N x\cos\theta_I}e^{i k_y y},
\end{equation}
using the following basis to describe the two conjugate frequencies:
\begin{equation}
    e^{-iE/\hbar t} \equiv \begin{pmatrix}1\\0\end{pmatrix};~
    e^{+iE/\hbar t} \equiv \begin{pmatrix}0\\1\end{pmatrix}.
\end{equation}
This unique incident wave provokes two different reflections. The first one is the normal reflection whose wave function reads:
\begin{equation}
    \psi_N = \begin{pmatrix}1\\0\end{pmatrix}e^{i k_N \cos\theta_N x}e^{i k_y y},
\end{equation}
and the second one is the Andreev-like reflection with a wave vector:
\begin{equation}
    k_A=\frac{\sqrt{2m(E_p-E)}}{\hbar}.
\end{equation}
This corresponds to a wave with reversed energy (with respect to the pump energy). Its wave function reads:
\begin{equation}
    \psi_A = \begin{pmatrix}0\\1\end{pmatrix} e^{i k_A \cos\theta_A x}e^{-i k_y y}.
\end{equation}
The relations between the different angles can be determined from the translational invariance of the problem along the $y$ direction, which yields~\cite{andreev1964thermal}:
\begin{equation}\label{transINV}
k_y=k_N\sin \theta_I=-k_N\sin \theta_N=k_A\sin \theta_A.
\end{equation}
This gives both the trivial result $\theta_N=-\theta_I$ and the nontrivial result
\begin{equation}
    \theta_A=\arcsin\left(\frac{k_N}{k_A}\sin \theta_I\right).
\end{equation}
This is analogous to the Snell-Descartes law, which has already been extended to interfaces of electronic systems~\cite{cheianov2007focusing,betancur2019electron}. However, there is no minus sign for the angle of the Andreev reflection, meaning that it occurs in the direction opposite to the incident wave.
The different angles considered $\theta_{I,N,A}$ are represented in real space in Fig.~\ref{fig2}(b) and in reciprocal space in Fig.~\ref{fig2}(c). The Andreev reflection occurs in the same quadrant as the incident wave, which is very different from the usual reflection, but close to the electronic Andreev reflection. In our case, however, the so-called Andreev approximation does not hold. In electronics, this approximation states that the Andreev-reflected particle has a direction strictly reversed with respect to the incident particle because the gap $\Delta$ is negligible with respect to the Fermi energy $\Delta\ll E_F$. In our case, the pump detuning $E_p$ plays the role of the Fermi energy $E_F\sim E_p$. The Andreev approximation in our case would consist in neglecting $\Delta$ with respect to $E_p$. However, in our system, the gap can be approximately equal to the pump detuning $\Delta \approx E_p$ ($2\Delta=E_p$ in most of the figures). 
Moreover, the expression of the angle clearly states that for incident particles of positive energies (still with respect to $E_p$), the Andreev-reflected particle will be reflected at a higher angle, giving upper bound for the angle of incidence $\theta_{I,c}$, above which no Andreev reflection is possible:
\begin{equation}\label{angleRc}
    \theta_{I,Rc}=\arcsin{\sqrt{\frac{E_p-E}{E_p+E}}}.
\end{equation}
Above this angle, the Andreev wave becomes a surface wave, evanescent on both sides of the interface.

\begin{figure}[tbp]
    \centering
    \includegraphics[width=0.99\linewidth]{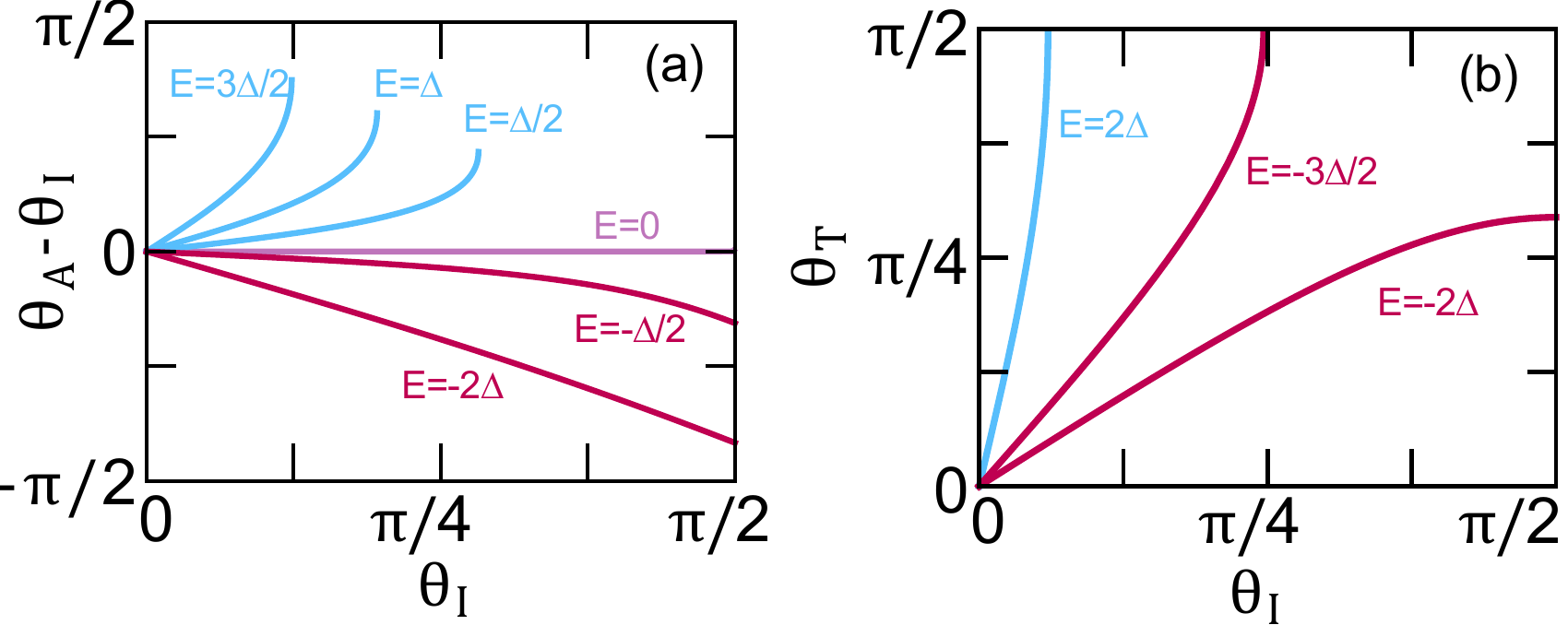}
    \caption{(a) Andreev reflection angle with respect to the angle of incidence for different energies of the incident wave. The difference is not negligible, because the Andreev approximation ($\Delta\ll E_p$) does not hold. (b) The angle of the transmitted wave with respect to the angle of incidence for different energies of the incident wave.}
    \label{fig3}
\end{figure}

On the contrary, if the energy of the incident particle is negative, the Andreev-reflected particle will be reflected at an angle smaller than the angle of incidence: $\theta_A\leq \theta_I$. There will be no limit angle in that case because for normal incidence $\theta_I=0$ the Andreev reflection will be normal as well. The dependence of the deviation of the angle of Andreev reflection from the angle of incidence $\theta_A-\theta_I$ on the angle of incidence $\theta_I$ for different energies is plotted in Fig.~\ref{fig3}(a). We can see that for positive energies, the curves exhibit a limit angle, whereas for negative energies there is no such bound, as stated above.

Now that the different propagative states are defined in the normal region, we can consider the scattering process of Andreev reflection and compute the scattering coefficients. In the case $E<\Delta$, there are no propagative states in the superfluids, so the wave function of the collective excitation in the superfluid~\eqref{wf} has to be written explicitly to account for the exponential spatial decay of the evanescent states:
\begin{widetext}
\begin{equation}\label{wfevan}
    \psi(x>0,y,t)=e^{-i\omega_p t}\left(\psi_s+ u e^{ik_y y}e^{-\kappa x} e^{-i\omega t}
    +v^* e^{-ik_y y}e^{-\kappa x}  e^{i\omega t}\right).
\end{equation}
\end{widetext}
We consider only real frequencies. $\kappa$ is the inverse decay length and $k_y$ is the wave vector in the $y$ direction, directly inherited from the incident wave. The propagative form of this term is responsible for the shift of finite-size beams upon reflection in the Goos-H\"anchen effect\cite{Goos1947} and its analogs.
Inserting the solution \eqref{wfevan} into the equation \eqref{GP} we find 
\begin{equation}\label{Ekevan}
    E^2=\left( \epsilon_{k_y}-\epsilon_{\kappa_x} + \alpha n -E_p\right)\left( \epsilon_{k_y}-\epsilon_{\kappa_x} + 3\alpha n -E_p\right).
\end{equation}
From this relation, we can find that there are two different inverse decay lengths at each energy, different from the spatially homogeneous case of  Eq.~\eqref{kkpm} because it takes into account the interface and the invariance along $y$:
\begin{equation}
    \kappa_\mp = \sqrt{k_y^2 + \frac{2m \left(2\alpha n-E_p\pm\sqrt{(\alpha n)^2+E^2} \right)} { \hbar^2}}.
\end{equation}

We define the scattering matrix in the $\{E,-E\}$ basis as:
\begin{equation}
    S=\begin{pmatrix}
    r_{N} & r_{A}^* \\
    r_{A} & r_{N}
\end{pmatrix},
\end{equation}
where $r_{N,A}$ are the scattering coefficients denoting normal/Andreev reflection, respectively. To determine them, we consider the wave functions in the normal region $\Psi_N$ and superfluid region $\Psi_S$ for a given configuration, say an incident wave from the normal region toward the interface at positive energy. This gives:
\begin{equation}
    \Psi_N=\frac{1}{\sqrt{w_+}}\left (\psi_I+r_N\psi_N\right)+\frac{1}{\sqrt{w_-}}\psi_A,
\end{equation}
where $w_\pm=\frac{\hbar}{m}k_{N,A} \cos\theta_I$ are the group velocities at $\pm E$, and
\begin{equation}
    \Psi_{S}=\sum_{*\equiv \pm}\eta_*e^{-\kappa_{*} x} \begin{pmatrix}u_*e^{i k_y y} \\v_*e^{-i k_y y}\end{pmatrix}
\end{equation}

The scattering coefficients are found from the continuity of the wave function and its derivative at the interface for both energies, which gives a system of equations:
\begin{equation}
       \left \{
   \begin{array}{r c l}
      \Psi_N(x=0) & = & \Psi_S(x=0) \\
      \frac{\partial \Psi_N}{\partial x}(x=0) & = & \frac{\partial \Psi_S}{\partial x}(x=0)
   \end{array}
   \right . ,
\end{equation}
which can be written in a complete form composed of 4 equations: 
\begin{widetext}
\begin{eqnarray}
    \frac{1}{\sqrt{w_+}}[1+r_N]e^{i k_y y} \begin{pmatrix}1\\0\end{pmatrix} 
    + \frac{r_A}{\sqrt{w_-}} e^{-i k_y y} \begin{pmatrix}0\\1\end{pmatrix}
    & = & 
    \eta_+\begin{pmatrix}u_+ e^{ik_y y}\\v_+ e^{-ik_y y}\end{pmatrix}
    + \eta_- \begin{pmatrix}u_- e^{ik_y y}\\v_- e^{-ik_y y}\end{pmatrix}, \\
    \frac{ik_N\cos{\theta_I}}{\sqrt{w_+}}[1-r_N]e^{i k_y y}\begin{pmatrix}1\\0\end{pmatrix}  
    - \frac{ik_A \cos{\theta_A} r_A}{\sqrt{w_-}}  e^{-i k_y y} \begin{pmatrix}0\\1\end{pmatrix}  & = &
    -\kappa_+ \eta_+\begin{pmatrix}u_+ e^{ik_y y} \\v_+e^{-ik_y y}\end{pmatrix}
    -\kappa_- \eta_- \begin{pmatrix}u_- e^{ik_y y}\\v_-e^{-ik_y y}\end{pmatrix}. 
\end{eqnarray}
The normal and Andreev reflection coefficients are then determined analytically:
\begin{eqnarray}
      r_N & = & \frac{(k_A \cos{\theta_A}+i\kappa_-)(k_N \cos{\theta_I}-i\kappa_+)u_+v_- - (k_N \cos{\theta_I}-i\kappa_-)(k_A \cos{\theta_A}+i\kappa_+)u_-v_+}{ (k_A \cos{\theta_A}+i\kappa_-)(k_N \cos{\theta_I}+i\kappa_+)u_+v_- - (k_N \cos{\theta_I}+i\kappa_-)(k_A \cos{\theta_A}+i\kappa_+)u_-v_+},\\
    r_A & = & \frac{2 i k_N \cos{\theta_I} (\kappa_+-\kappa_-) v_- v_+ }{ (k_A \cos{\theta_A}+i\kappa_-)(k_N \cos{\theta_I}+i\kappa_+)u_+v_- - (k_N \cos{\theta_I}+i\kappa_-)(k_A \cos{\theta_A}+i\kappa_+)u_-v_+} \frac{\sqrt{w_-}}{\sqrt{w_+}}.
\end{eqnarray}
\end{widetext}

\begin{figure}[tbp]
    \centering
    \includegraphics[width=0.99\linewidth]{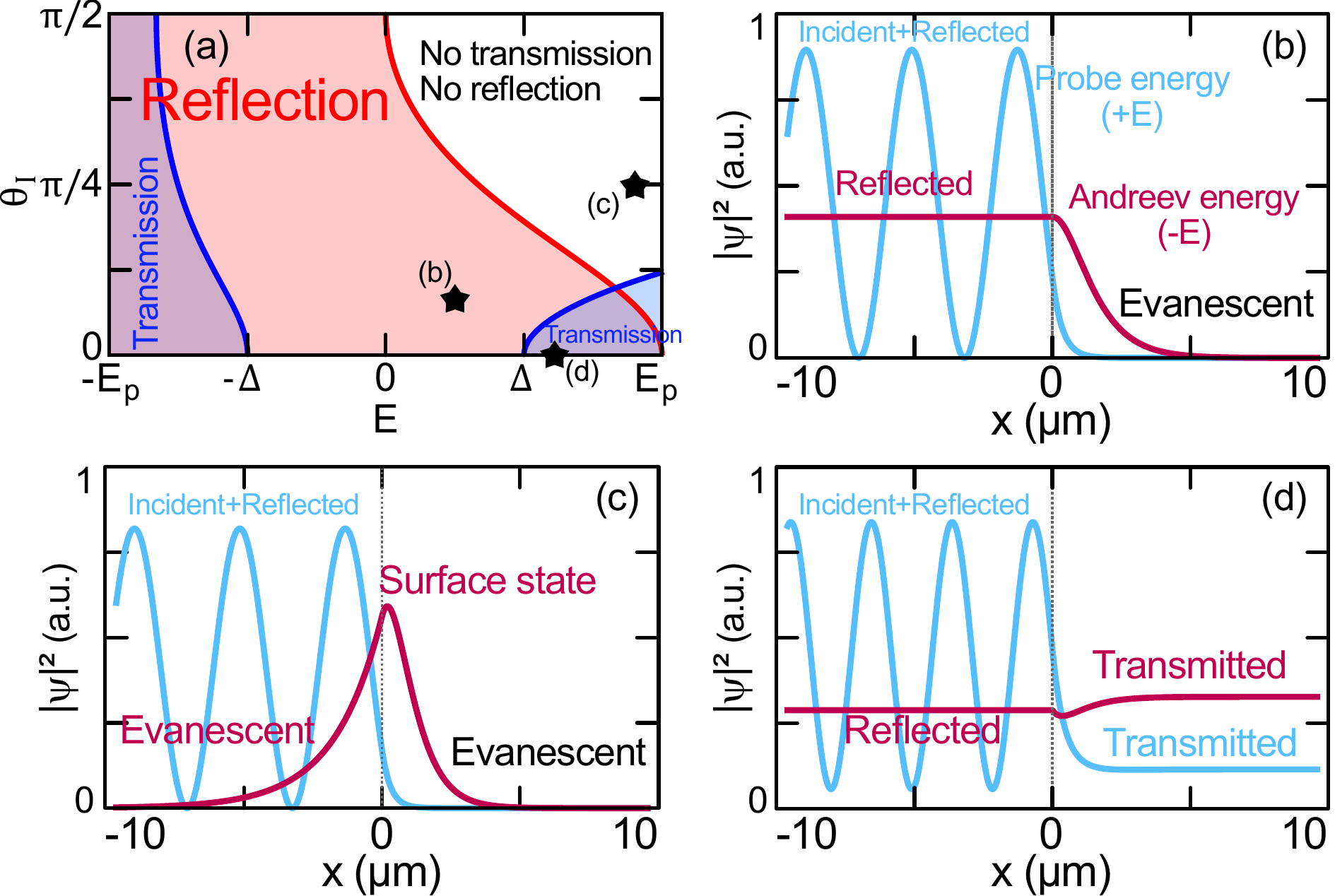}
    \caption{(a) Phase diagram showing the experimental conditions $E,\,\theta_I$ leading to Andreev reflection (red) and/or transmission (blue areas). The domains are limited by red/blue thick lines respectively. (b-d) Real space profile of the states at positive (blue) and negative (red) energy for three different experimental conditions. (b) $\theta_I=\pi/6$, $E=\Delta/2=0.125\,$meV leads to Andreev reflection but no transmission, all states are evanescent in the superfluid. (c) $\theta_I=\pi/4$, $E=0.45\,$meV, where there are only evanescent states at the negative energy, which creates a surface state. (d) $\theta_I=0$, $E=0.3\,$meV, which give both an Andreev reflection and transmission in the superfluid. Note that specular reflection is present in all cases, which leads to interferences (with the incident probe) in the amplitude at positive energy in the normal region.}
    \label{fig4}
\end{figure}

In the end, all the quantities can be calculated from the experimental conditions. As shown in Fig.~\ref{fig2}(d), a unique incident wave gives indeed two different reflections (for $\theta_I<\theta_{I,Rc}$). The first one (specular reflection) is at the energy of the incident wave, with an amplitude $r_N$. The second one (Andreev reflection) is at the conjugate frequency, with an amplitude $r_A$. There is no transmission in the superfluid, only evanescent states. The profile of the corresponding wave function in real space is plotted in Fig.~\ref{fig4}(b). The incident and reflected waves at the incident frequency (blue) interfere, leading to a spatial modulation of the wave function at positive energy on the left-hand side of the interface ($x<0$, normal region). On the contrary, there is only one reflected wave at the Andreev frequency (in red), which gives a constant profile in the normal region. In the superfluid region, at both energies, the states are evanescent and the norm of the wave functions rapidly decreases to zero. Above the limit angle, the Andreev reflection coefficient goes to zero, meaning that Andreev reflection cannot occur anymore, because the wave vector $k_y$ dictated by experimental conditions is larger than $k_A$ (see Fig\ref{fig2}(c), in the case of the black dotted line). If Andreev reflection is impossible, the profile of the wave functions will look as shown in Fig.~\ref{fig4}(c). At positive energy, this is very similar to the case of Fig.~\ref{fig4}(b), but at negative energy, Andreev reflected waves cannot propagate in the normal region and are evanescent, as well as in the superfluid. In the end, the state at negative energy is evanescent at both sides of the interface, so it is a surface state exponentially localized at the interface.

\subsection{$E>\Delta$}
Previously, we considered only the case $E<\Delta$, that is, energies lying inside the gap. However, Andreev reflection can occur even outside the gap. The derivation is very similar to what was shown previously. However, one should work with the propagative states in the superfluid, with the wave vectors~\eqref{kpm} (we remind that even for $E>\Delta$, there are still evanescent states in the superfluid in addition to the propagative states, describing the wave profile close to the interface). The particle is transmitted at an angle of:
\begin{equation}
    \theta_T=\arcsin \left( \frac{k_T}{k_N}\sin \theta_I\right),
\end{equation}
determined by $k_y=k_N \sin \theta_I = k_T \sin \theta_T$, where $k_T\equiv k_+$ is the wave vector of the transmitted particle. Since $k_T>k_N$, there is a critical angle of incidence above which there is no transmission, even for $E>\Delta$:
\begin{equation}\label{angleTc}
    \theta_{I,Tc}=\arcsin \left (\frac{E_p - 2 \alpha n + \sqrt{(\alpha n)^2+E^2}}{E_p+E}\right).
\end{equation}
Figure~\ref{fig3}(b) shows the angle of transmission with respect to the angle of incidence for different energies. Again, there is a critical angle for certain energies, when $\theta_T$ reaches the value $\pi/2$, as shown in this figure.

We represented the domains of energy/angle of incident particle where Andreev reflection (red) or transmission (blue) is possible in \ref{fig4}(a) using Eqs~\ref{angleRc} and~\ref{angleTc}. We see that the effective gap $\Tilde{\Delta}$ where no transmission is possible is equal to the gap $\Delta$ calculated in Eq.~\eqref{gap} only for $\theta_I=0$. Away from normal incidence, the effective gap is much larger than this value, meaning that the transmission is possible only for small angles of incidence.

\begin{figure}[tbp]
    \centering
    \includegraphics[width=0.99\linewidth]{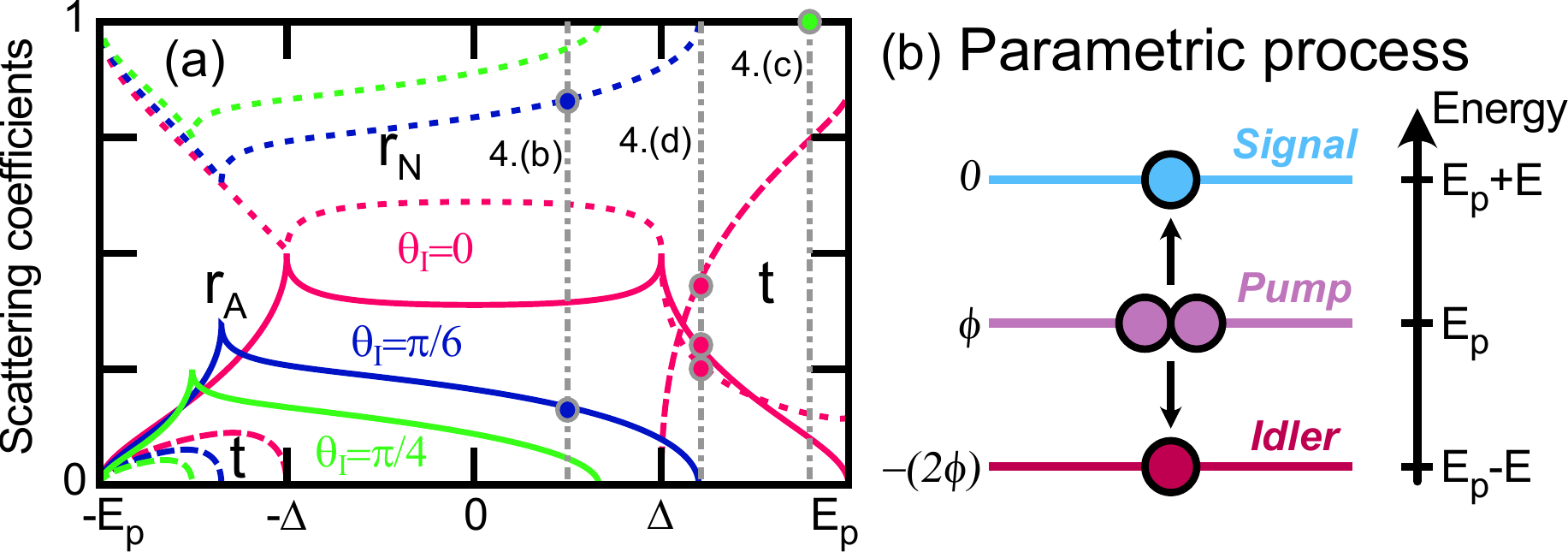}
    \caption{(a) Scattering coefficients $r_N$ (dotted), $r_A$ (solid) and $t$ (dashed lines) with respect to the energy for different angles of incidence $\theta_I=\{0,\pi/6,\pi/4\}$ in pink, blue and green respectively. The grey lines (and corresponding points) show the energy (and corresponding angles) for which are calculated the states in Fig.~\ref{fig4}. Note the peaks in the remaining coefficients when one vanishes. (b) Illustration of the parametric process in Andreev reflection. The pump is at energy $E_p$ with phase $\phi$. Positive energies are at $E_p+E$ with no phase (we disregard the phase of the incident wave) whereas negative energies at $E_p-E$ gain a phase $-2\phi$ because of the parametric process (the minus sign appears because of the conjugation, see Eq.~\eqref{wf}). 2 "particles" from the pump are converted in one "particle" at $+E$ and no phase and one "particle" at $-E$ with a phase $-2\phi$ according to the phase-matching condition~\eqref{phase}.}
    \label{fig5}
\end{figure}

Moreover, in the region $E>\Delta$ the scattering coefficients change with respect to those computed for $E<\Delta$, and there is one more scattering coefficient to determine, the one accounting for the transmission in the superfluid. We can write the wave function in the superfluid in this case:
\begin{equation}\label{psist}
    \Psi_{S,t}=t \begin{pmatrix}u_*e^{i (k_+ x +k_y y)} \\v_*e^{-i (k_+x +k_y y)}\end{pmatrix}+\eta_-e^{-\kappa_{-} x} \begin{pmatrix}u_*e^{i k_y y} \\v_*e^{-i k_y y}\end{pmatrix}
\end{equation}
and we finally find the transmission coefficient by considering the continuity of the wave function and its derivative at the interface:
\begin{widetext}
\begin{eqnarray}
    r_N & = & \frac{(k_A \cos{\theta_A}+i\kappa_-)(k_N \cos{\theta_I}-k_T\cos\theta_T)u_+v_- - (k_N \cos{\theta_I}-i\kappa_-)(k_A \cos{\theta_A}-k_T\cos\theta_T)u_-v_+}{(k_A \cos{\theta_A}+i\kappa_-)(k_N \cos{\theta_I}+k_T\cos\theta_T)u_+v_- - (k_N \cos{\theta_I}+i\kappa_-)(k_A \cos{\theta_A}-k_T\cos\theta_T)u_-v_+},\\
    r_A & = & \frac{2k_N \cos \theta_I (k_T\cos \theta_T+i\kappa_-)v_-v_+}{(k_A \cos{\theta_A}+i\kappa_-)(k_N \cos{\theta_I}+k_T\cos\theta_T)u_+v_- - (k_N \cos{\theta_I}+i\kappa_-)(k_A \cos{\theta_A}-k_T\cos\theta_T)u_-v_+} \frac{\sqrt{w_-}}{\sqrt{w_+}},\\
    t & = & \frac{2k_N \cos \theta_I (k_A\cos \theta_A+i\kappa_-)v_-}{(k_A \cos{\theta_A}+i\kappa_-)(k_N \cos{\theta_I}+k_T\cos\theta_T)u_+v_- - (k_N \cos{\theta_I}+i\kappa_-)(k_A \cos{\theta_A}-k_T\cos\theta_T)u_-v_+} \frac{\sqrt{w_T}}{\sqrt{w_+}}
\end{eqnarray}
where $w_T=\xi\hbar k_T \cos\theta_T/m $ is the group velocity for waves transmitted in the superfluid, with $\xi=(2\alpha n-E_p+\epsilon_k)/E$. Using $u_+=-v_-e^{2i\phi}$ and $v_+=u_-e^{2i\phi}$, we can rewrite these expressions as:
\begin{eqnarray}
    r_N & = & \frac{(k_A \cos{\theta_A}+i\kappa_-)(k_N \cos{\theta_I}-k_T\cos\theta_T)v^2 + (k_N \cos{\theta_I}-i\kappa_-)(k_A \cos{\theta_A}-k_T\cos\theta_T)u^2}{(k_A \cos{\theta_A}+i\kappa_-)(k_N \cos{\theta_I}+k_T\cos\theta_T)v^2 + (k_N \cos{\theta_I}+i\kappa_-)(k_A \cos{\theta_A}-k_T\cos\theta_T)u^2},\\
    r_A & = & \frac{2\sqrt{k_N k_A} \cos \theta_I (k_T\cos \theta_T+i\kappa_-)uv}{(k_A \cos{\theta_A}+i\kappa_-)(k_N \cos{\theta_I}+k_T\cos\theta_T)v^2 + (k_N \cos{\theta_I}+i\kappa_-)(k_A \cos{\theta_A}-k_T\cos\theta_T)u^2}e^{-2i\phi} ,\\
    t & = & \frac{2\sqrt{k_N k_T} \cos \theta_I (k_A\cos \theta_A+i\kappa_-)v}{(k_A \cos{\theta_A}+i\kappa_-)(k_N \cos{\theta_I}+k_T\cos\theta_T)v^2 + (k_N \cos{\theta_I}+i\kappa_-)(k_A \cos{\theta_A}-k_T\cos\theta_T)u^2}e^{-i\phi} 
\end{eqnarray}
\end{widetext}
with $u=|u_-|$ and $v=|v_-|$. We will explain a few features of these scattering coefficients. First, one can verify that:
\begin{equation}
    |r_N|^2+|r_A|^2\mathcal{U}(\theta_{I,Rc}-\theta_I)+|t|^2 \mathcal{U}(|E|-\Tilde{\Delta})=1,
\end{equation}
where $\mathcal{U}$ denotes the Heaviside step function. This equation accounts for the conservation of the current density: an incident wave of amplitude 1 is partly reflected at the same frequency with amplitude $|r_N|^2$, partly at the conjugate frequency with amplitude $|r_A|^2$ (if the angle of incidence $\theta_I<\theta_{I,Rc}$), and partly transmitted with amplitude $|t|^2$ (if the energy $|E|>\Tilde{\Delta}$) by a wave combining both frequencies. Moreover, we can see that the phase of the superfluid plays a role only in the components at $-E$. Indeed, it does not appear in $r_N$, but only in $r_A$ (with a factor 2) and in $t$. The wave function transmitted in the superfluid is proportional to $t u$ at $E$ and $t v$ at $-E$ (see Fig.~\ref{fig2}(d)), which means that the total phase of the transmitted wave is 0 at $E$ and $-2\phi$ at $-E$, in agreement with the phase matching condition of the parametric process discussed below and shown in Fig~\ref{fig5}(b). 

We already discussed that there are experimental conditions leading to the presence of Andreev reflection and the absence of transmission (Fig.~\ref{fig4}(b)), or to the absence of both Andreev reflection and transmission, accompanied by the formation of a surface wave (Fig.~\ref{fig4}(c)). As shown above, there are also experimental conditions where the transmission (and eventually Andreev reflection) can be observed. The profile of the state in this case is shown in Fig.~\ref{fig4}(d): there is a transmission in the superfluid, and simultaneously evanescent states in the superfluid, as stated previously, whose influence on the profile is visible only close to the interface. Far from the interface, there is one plane wave at each energy, as expected from the form of the wave function~\eqref{psist}. This is the propagative bogolon quasiparticle, with a part at positive and a part at negative energy, whose respective amplitudes are determined by $u$ and $v$.

Fig.~\ref{fig5}(a) shows the variation of the scattering coefficients $r_N$, $r_A$, and $t$ for an incident wave of energy $-E_p<E<E_p$ and with an angle of incidence $\theta_I$ (different angles are plotted with different colors). Increasing the angle of incidence generally decreases the Andreev reflection and transmission coefficients (the normal reflection thus increases, because the sum of all square norms of the scattering coefficients always equals $1$). At $\theta_I=\pi/4$, $r_A$ cancels for some frequency range within the gap. Here the mode at Andreev frequency is evanescent both in the superfluid and normal region, forming a propagating surface wave. It is clear that the particle-hole symmetry (mapping from $E$ to $-E$) only holds for $|E|<\Delta$ and at normal incidence, so it is a very fragile case. Moreover, some coefficients exhibit infinite derivatives, which correspond to the appearance or disappearance of another coefficient. For instance,  for $\theta_I=0$ and $E=\Delta$, there is a peak with an infinite derivative for both the normal and Andreev reflections, $r_N$ and $r_A$. For $E\to\Delta^-$, the inverse decay length $\kappa_- \to 0$. Thus, the associated wave is less and less vanishing, which enhances the probability of energy conversion to occur, thus increasing the Andreev reflection (and decreasing the normal reflection at the same time). Reversely, for $E\to\Delta^+$, the wavevector $k_T \to 0$, and the wave propagates critically slowly in the superfluid (we remind that the dispersion of the gapped superfluid is parabolic), which again enhances the probability of energy conversion. Finally, for large energies $E\ll\Delta$, the transmission in the superfluid region becomes dominant over reflection.

In Fig.~\ref{fig5} we also indicate the regimes corresponding to the previous figures. First, we have considered the case of a wave incident at positive energy $E=1.2\Delta$, and at normal incidence $\theta_I=0$. This corresponds to the vertical grey line at this energy and the pink curves in Fig.~\ref{fig5}(a). The grey line cuts the three pink curves when none of them is zero, meaning that there is a specular reflection $r_N\neq 0$ (dotted line), an Andreev reflection $r_A\neq 0$ (solid line), and a transmission $t\neq 0$ (dashed line). This is consistent with the phase diagram of Fig.~\ref{fig4}(a), which states that there is both transmission and Andreev reflection in this region (specular reflection is always possible). The state found in this case is plotted in Fig.~\ref{fig4}(d), and the Andreev reflected wave is visible, as well as the transmitted bogolons. 

The second case corresponds to an incident wave of positive energy $E=1.8\Delta$, and an angle of incidence of $\theta_I=\pi/4$ (the set of green curves). The vertical grey line at this energy cuts only the normal reflection coefficient, whose norm is $1$ because Andreev reflection and transmission coefficients are zero at this energy (for instance, the Andreev reflection coefficient decreases to $0$ shortly after $E=\Delta/2$). For those experimental conditions, the phase diagram indeed tells us that there is no Andreev reflection nor transmission possible. The profile of this state is shown in Fig.~\ref{fig4}(c). The wave at the Andreev frequency is evanescent in both regions, so the state is localized at the interface.

Finally, the third state corresponds to an incident energy $E=\Delta/2$ (within the gap) and an angle of incidence of $\theta_I=\pi/6$. The vertical grey line cuts only two blue lines: $r_N\neq 0$ and $r_A\neq 0$. There is no transmission ($E<\Delta$). The resulting state is shown in Fig.~\ref{fig4}(b).

Note that we completely disregard the case of a wave coming from the superfluid towards the interface, because we believe that the corresponding experiment is much more difficult to perform in a controlled way, creating a well-defined probe with $u$ and $v$.

We note once again that a phase $2\phi$ appears in the wave functions at the Andreev frequency (both in the reflected and transmitted wave) whereas there is no superfluid phase associated with the wave function at the pump frequency. This is known to occur in optical parametric amplification, as depicted in Fig.~\ref{fig5}(b). The superfluid plays the role of a pump, the normal frequency is the source, and the Andreev frequency is the idler. From two pump particles, one gets one particle at each energy $E_p\pm E$, which verifies the energy and momentum conservation laws:
\begin{equation}\label{phase}
    \begin{array}{r c l}
      2 E_p & = & (E_p+E) + (E_p-E) \\
      2 i\phi & = & 0 +((- 2\phi)i)^*,
   \end{array}
\end{equation}
where the conjugation appears from Eq.~\eqref{wf}.

\section{Simulations}
\begin{figure}[tbp]
    \centering
    \includegraphics[width=0.99\linewidth]{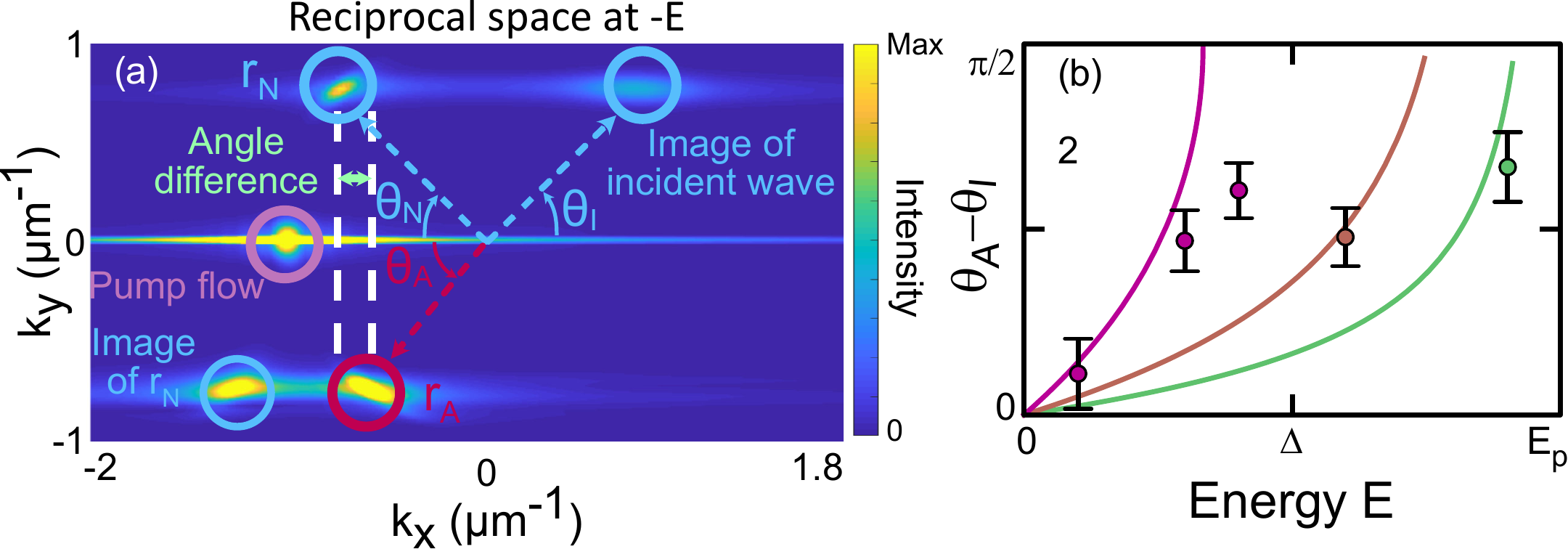}
    \caption{(a) The intensity distribution in the reciprocal space (measured from the N-region) at the "Andreev" frequency ($-E$ for an incident probe at $+E$) in a realistic configuration where interactions are also present in the "normal" region. The angles are noted directly on the figure, as done schematically in Fig.~\ref{fig2}(c). The difference in $k_x$ for normal and Andreev reflections is visible. (b) Difference between the angle of incidence and the angle of Andreev reflection with respect to the energy for $\theta_I=\{45^{\circ},20^{\circ},10^{\circ}\}$ (resp. purple/brown/green). Lines represent theory and points results from numerical experiments.}
    \label{fig6}
\end{figure}

To confirm that the reflection can be observed experimentally, we simulated the Gross-Pitaevskii equation~\eqref{GP} numerically. In addition to the theoretical model, we added interactions in the normal region, as expected in an experiment (however, the normal region is, of course, not pumped). The numerical experiment consists in sending a wave at a given energy $E-E_p=0.4\Delta=0.1$~meV and given angle of incidence ($\theta_I=45^{\circ}$) from the normal region to the superfluid. The parameters of the simulation are: $m=8\times 10^{-5}m_0$ ($m_0$ is the free electron mass), $\gamma=\hbar/2\tau$ with $\tau=15$~ps, $\alpha=3.6$~$\mu$eV$\cdot \mu$m$^2$, $E_p=0.5$~meV.
The probe needs to be sent from a reasonably short distance (15~$\mu$m), because of the polariton decay. The superfluid created by the resonant pump expands partly in the normal region. Fig.~\ref{fig6}(a) shows the distribution of the probability density in the reciprocal space at the Andreev frequency (conjugate to the frequency of the incident wave). We can see that the interactions in the normal region make appear several different intensity peaks not expected in the ideal picture analytically considered. First, the pump at the energy $E_p$ and $r_N$ at the energy $E$ are still visible at $-E$ because of their finite frequency width in a finite-time simulation. One can also see one peak corresponding to the "image" of the incident wave and another which is the image of the normally-reflected wave. These both images are created by the non-linear parametric process which now couples $E$ and $-E$ in the normal region. However, the most important peak is the Andreev reflection, clearly visible and well-separated from all other signals in k-space. 

In addition, the direction of the Andreev reflection is not exactly opposite to that of the incident wave, as expected ($\theta_A\neq\theta_I)$.
We quantitatively verify the agreement between theoretical and numerical results in Fig.~\ref{fig6}(b). The difference between the angle of Andreev reflection and the angle of incidence is plotted with respect to the energy for different angles of incidence. We can see that there is a good agreement between theoretical (lines) and numerical (points with error bars) results, even if the latter were performed in a realistic case (finite lifetime, pump flow in the normal region, etc). 


\section{Conclusion}
To conclude, we provide a comprehensive theoretical analysis of the angular-dependent Andreev reflection on a polaritonic gapped superfluid. The regime of gapped superfluid leads to evanescent states in the superfluid, whereas the non-superfluid region can have propagative states. It enables us to observe a phenomenon analogous to the Andreev reflection: a wave incident at $+E$ is reflected at $-E$. The Andreev-reflected wave is reflected at an angle different from the angle of incidence. We find a critical angle, above which the Andreev reflection cancels, and the Andreev wave becomes a surface mode. We find a good agreement between our analytical model and realistic numerical simulations. The phenomenon we describe is close to optical phase conjugation and analogous to reflection on black hole horizons.

\begin{acknowledgments}
We thank J. Meyer, A. Bramati, and D. Sanvitto for inspiring discussions.
This research was supported by the ANR Labex GaNext (ANR-11-LABX-0014), the ANR program "Investissements d'Avenir" through the IDEX-ISITE initiative 16-IDEX-0001 (CAP 20-25), the ANR project "NEWAVE" (ANR-21-CE24-0019) and the European Union's Horizon 2020 program, through a FET Open research and innovation action under the grant agreement No. 964770 (TopoLight).
\end{acknowledgments}

\newpage

\bibliography{biblio} 
\end{document}